# Performance Analysis of DNA Crossbar Arrays for High-Density Memory Storage Applications

Arpan De[1], Hashem Mohammad[1,2], Yiren Wang[1], Rajkumar Kubendran[3], Arindam K. Das[4], M.P.Anantram[1,**]

[1]Department of Electrical and Computer Engineering, University of Washington, 98195 Seattle, WA, USA
[2]Department of Electrical Engineering, Kuwait University, P.O. Box 5969, Safat 13060, Kuwait
[3]Department of Electrical and Computer Engineering, University of Pittsburgh, 15261 Pittsburgh, PA, USA
[4]Department of Computer Science and Electrical Engineering, Eastern Washington University, 99004 Cheney, WA, USA.

** Corresponding author: M. P. Anantram (e-mail: anant@uw.edu)

## ABSTRACT

Deoxyribonucleic acid (DNA) has emerged as a promising building block for next-generation ultra-high density storage devices. Although DNA has high durability and extremely high density in nature, its potential as the basis of storage devices is currently hindered by limitations such as expensive and complex fabrication processes and time-consuming read-write operations. In this article, we propose the use of a DNA crossbar array architecture for an electrically readable Read-Only Memory (DNA-ROM). While information can be 'written' error-free to a DNA-ROM array using appropriate sequence encoding, its read accuracy can be affected by several factors such as array size, interconnect resistance, and Fermi energy deviations from HOMO levels of DNA strands employed in the crossbar. We study the impact of array size and interconnect resistance on the bit error rate of a DNA-ROM array through extensive Monte Carlo simulations. We have also analyzed the performance of our proposed DNA crossbar array for an image storage application, as a function of array size and interconnect resistance. While we expect that future advances in bioengineering and materials science will address some of the fabrication challenges associated with DNA crossbar arrays, we believe that the comprehensive body of results we present in this paper establishes the technical viability of DNA crossbar arrays as low power, high-density storage devices. Finally, our analysis of array performance vis-à-vis interconnect resistance should provide valuable insights into aspects of the fabrication process such as proper choice of interconnects necessary for ensuring high read accuracies.

## 1. INTRODUCTION

Production of digital data is increasing exponentially, driven predominantly by the adoption of artificial intelligence (AI) and deep learning. According to current projections, approximately 175 zettabytes (109 terabytes) of new data will be generated annually by 2025. Archival storage of even a fraction of such immense data requires a fundamental rethink in durable, low power, accurate, and extreme high-density storage technology since traditional storage devices (e.g., magnetic tapes, hard disks, Blu-ray, and solid-state devices like flash) are fast approaching their scaling and performance limits. The key figures-of-merit of a storage technology are: (i) the maximum density of any storage device in terms of bits per unit volume, (ii) durability (total time for which the data can be stored), and (iii) energy cost. With the projected explosion of data storage requirements, it is widely accepted that existing technologies are inadequate with respect to the figures-of-merit. Silicon-based devices are fast approaching their density limits and further downscaling of such devices will result in performance degradation. This motivates the demand for development of alternate ultra-high-density storage technologies which are durable and energy-efficient.

Deoxyribonucleic acid (DNA) is one of the most promising candidates for high-density storage devices. The concept of using DNA for data storage dates to the mid-1960s when the idea of 'genetic memory' was proposed by Neiman and Wiener [1], [2]. However, technological limitations at the time regarding DNA sequencing and synthesis limited the practical viability of such memory devices. Around 1986, Davis et al [3] and Clelland et al [4] made the first experimental demonstrations of DNA storage. A breakthrough was achieved in 2012-13 by successfully storing hundreds of kilobytes of data [5].

DNA-based molecular-level data storage is an excellent choice for future high-density storage applications since only a few layers of atoms are required to store one bit of information. DNA storage can be significantly denser than the densest storage device available at present and is also more durable. When insulated from heat, humidity, and light, DNA can last for centuries to millennia, compared to decades for traditional media devices. Table 1 provides a comparison of conventional silicon-based storage devices, resistive random-access memory (RRAM), and DNA-based devices with respect to (w.r.t) density, random access, read latency, and durability.

DNA based data storage involves four steps: (i) encoding (mapping of digital information to appropriate DNA sequences), (ii) synthesis (writing into DNA sequences), (iii) reading (accessing the data), and (iv) decoding (converting to digital format). Current DNA-in-solution storage systems are engineered to store data in chunks of DNA [6], [7]. Traditionally, information in DNA-in-solution systems is encoded into appropriate DNA sequences using a dedicated algorithm that maps digital bits to DNA bases [8]. Different methodologies for DNA sequence synthesis exist, for instance, phosphoramidite-based synthesis or enzymatic synthesis [9]. Polymerase chain reaction (PCR) is a proven high-throughput method that can facilitate large-scale DNA replication. The read process for DNA-in-solution systems has significant time complexity since it requires an indexing system for proper reconstruction of the data.

A major drawback of DNA-in-solution storage systems concerns random access. Even if a small segment of the stored data needs to be accessed and retrieved, the entire DNA pool must be read and sequenced, which is time-consuming and inefficient. However, some recent work exists that aims to address/improve random access, e.g., the magnetic bead extraction approach or PCR with primers discussed in [10]. We note that the read accuracy of DNA-in-solution storage systems depends largely on the synthesis process, which, as mentioned previously, is susceptible to errors.

Even though DNA-based storage systems have a transformative potential for next-generation storage applications, the technology is still at a nascent stage . In this paper, we address the random access and read latency issues by adopting a crossbar array architecture for an electrically readable Read-Only Memory (ROM) device [11]. Our proposed device is composed of interconnects and two different DNA sequences serving as memory cells, as illustrated in Figure 1. In contrast to existing DNA-in-solution storage systems where bit strings are encoded into a DNA sequence, we choose an 'appropriate' pair of DNA sequences and exploit the difference between their conductance to reliably store binary information. Motivated by recent advances in the field of synthetic biology, we propose self-assembly of crossbar nanostructures using DNA origami-inspired approaches [12]. The assembly of DNA crossbars will consist of separately forming the memory elements and the DNA nanowires, followed by their integration. Finally, after the self-assembly phase, the DNA crossbar will be connected to electrodes at the edges of the array to enable readout [11]. From a fabrication perspective, we propose leveraging synthetic biology approaches for fabricating and programming DNA memory cells/interconnects and conventional semiconductor and lithography processes for contact formation. Challenges related to DNA crossbar array fabrication include: (i) High precision contact formation between DNA memory cells and DNA nanowires (interconnects). Since DNA crossbar formation is a self-assembly process, a suitable fabrication recipe is needed which will yield good contacts between the memory cells and the nanowires. (ii) Fabrication of highly conductive DNA nanowires, an area which is mostly unexplored. The work by Ke and Hihath is a pioneering effort that has just stared to experimentally address this issue [13]. As seen from our computational analysis, interconnects should be conductive to achieve good read-out accuracy. Since DNA nanowires are typically very resistive, making them more conductive is a challenging part of the process. However, the relative contributions of the contact and intrinsic resistance of the DNA are unclear at this stage. Finally, as noted in [14], initialization of large DNA crossbar arrays can incur a large spatiotemporal overhead. Exceptional spatial control of self-assembly has been demonstrated for DNA origami [15], [16]. From a temporal perspective, self-assembly process can take up to 1-3 days [17], [18]. We estimate that self-origami of a DNA array can take up to 2-4 days, but the process is extremely parallelized such that billions of copies can be made in parallel.`

Exponential growth in data generation and the need for durable and long-term data storage have exposed the shortcomings of current memory technologies such as flash, RRAM, phase change memory (PCM), and magnetic tunnel junctions (MTJ). According to the International Roadmap for Devices and Systems (IRDS 2020), DNA storage technologies are a promising candidate for long-term, large-scale data storage applications. The development of commercially viable DNA-based storage devices is now widely recognized as a critical emergent research area and has rightfully attracted a proliferation of research in recent times. Nevertheless, the state-of-the-art in this area is still quite nascent and the path from ideation to commercialization is riddled with significant engineering and scientific challenges. Overcoming these challenges will require potentially ground-breaking innovations from researchers in diverse disciplines such as molecular biology, materials science, and electrical engineering. Some notable research challenges include making conducting DNA origami and the development of robust biotic-abiotic interfaces between DNA and the electrode array. Recent research trends in the domain of CMOS-DNA systems [19], [20] may provide a solution for integrating silicon-based devices/circuits with DNA. Equally important are the design of peripheral circuitry and the choice of interconnect material for nanoscale devices.

Despite the advantages of a crossbar topology over DNA-in-solution systems, two factors can adversely affect its read accuracy. First, Fermi energy variations [21] can alter the conductance state of DNA, leading to errors during the read-out process. Second, sneak paths, which distort the read current distributions in memory cells, affect all crossbar topologies and DNA-based arrays are no exception. Existing approaches for alleviating sneak path issues include the use of a selector switch coupled with an appropriate biasing scheme. While the ideal mechanism for mitigating sneak path effects in a DNA crossbar array is an open question, the feasibility of 'DNA diodes' [22] is an interesting proposition and merits further investigation. Additionally, the development of the read-out circuitry for DNA crossbar arrays offers significant engineering challenges and is subject to further research.

The rest of the paper is structured as follows. The calculation of current-voltage characteristics is addressed in Section 2, which is followed by a discussion of architectural concepts of a DNA-ROM array in Section 3. In Section 4, we study the performance of the array for different DNA strands and interconnect resistances. The impact of Fermi energy variation on array performance is addressed in Section 5. In Section 6, we present a detailed scalability analysis in terms of two key metrics – bit error rate and power consumption. We conclude this paper with a brief discussion of ongoing and future work in Section 7.

## 2. DNA CURRENT SIMULATION FRAMEWORK & RESULTS

We adopted two B-form DNA sequences 3'-CCCTCCC-5' and 3'-TTTCTTT-5, referred to as B-CT$_1$C and B-TC$_1$T respectively in this paper, which are studied in [23]. The simulation methodology involves four main steps: structure building, energy minimization, density functional theory calculations, and charge transport calculations. First, the Nucleic Acid Builder (NAB) software package [24] is used to obtain the atomic coordinates of double-stranded B-DNA structures. Next, we used PACMOL [25] to obtain energy minimized structures with explicit solvent (water) environment within a 60 Å x 60 Å x 60 Å cubic solvation box. We added Na counterions to neutralize the phosphate groups in the DNA backbone. We also added Na and Cl ions randomly into the solution to represent a homogenous mixture of 150 mM [26], [27]. We used General Amber Force Field [28]. For the water solvent, a truncated octahedron box of SPC/E water model was used to solvate the DNA. Amber force field with parmbsc1 correction [29] was used to energy minimize the structures. The structure minimization was done in two stages. The first stage consists of restraining the DNA with a force of 25 kcal/mol-Å2 for the first 1000 cycles, while the solvent environment is minimized. The second stage involves minimizing the whole system for additional 1000 cycles. Both stages of minimization have 500 steps of the steepest descent method, followed by 500 steps of the conjugate gradient method.

Next, we delete the solvent atoms and only keep the DNA atoms along with the closest Na counterions necessary to neutralize the DNA molecule. We then performed density functional theory calculations and obtain overlap and Fock matrices, denoted by S and F respectively. In this step, we used the B3LYP functional and 6-31G(d,p) basis set. We used the polarizable continuum model (PCM) to include the impact of water solvent environment. The Fock and overlap matrices resulting from this step are used as inputs for the charge transport calculations. The system Hamiltonian is obtained by transforming the F and S matrices to an orthogonal basis by applying the Lőewdin transformation [30]:

$$H_a = S^{-\frac{1}{2}} F S^{-\frac{1}{2}} \qquad (1)$$

where $H_a$ is the system Hamiltonian. The diagonal elements of $H_a$ represent the atomic orbitals and the off-diagonal elements represent the electronic coupling between these atomic orbitals. We chose to partition the DNA into nucleotides (backbone + base). We achieved this by transforming the Hamiltonian into a block-diagonal matrix, through the unitary transformation:

$$H_b = U^\dagger H_a U \qquad (2)$$

where U consists of the eigenvectors of each nucleotide. The eigenvectors in U are arranged according to the order of the DNA nucleotides. As a result, the diagonal elements of each diagonal block in $H_b$ represent the molecular orbital energies of the nucleotide and the off-diagonal blocks in $H_b$ represent the electronic coupling between the different nucleotides. Both dimensions of a diagonal block for $H_a$ and $H_b$ are equal to the number of orbitals used to represent the DNA nucleotide. We also note here that in obtaining $H_b$, we omitted the atomic orbital contributions of the counterions and focused the charge transport calculations on the DNA molecule itself. This is in accordance with the finding in [23] that the counterions do not impact the DNA conductance, especially in energy ranges within the vicinity of the highest occupied molecular orbital (HOMO) of the DNA.

After the $H_b$ matrix is obtained, the transmission versus energy of a DNA molecule is computed using the Green's function approach to account for decoherence [30]. We use this method to model the system shown in Figure 2(a). A DNA strand is placed between two metallic contacts and a voltage bias is applied. The governing Green's function equation is defined as:

$$[E - (H_b + \Sigma_L + \Sigma_R + \Sigma_B)]G^r = I \qquad (3)$$

where $E$ is the energy, $\Sigma_{L(R)}$ is the left (right) contact self-energy, $\Sigma_B$ is the decoherence self-energy, $G^r$ is the retarded Green's function and I is the identity matrix. In this formalism, we use the contact's self-energy matrices ($\Sigma_{L(R)}$) to account for the two electrodes shown in Figure 2(a). The self-energy of the contacts is defined as $\Sigma_{L(R)} = -i\Gamma_{L(R)}/2$, $i = \sqrt{-1}$ and $\Gamma_{L(R)} = 1000$ meV is the coupling value. The contact self-energies are added to the 3' and 5' nucleotide ends of the strand, respectively. The decoherence probe's self-energy is defined in a similar manner, $\Sigma_B = \sum_k(-\frac{i\Gamma_k}{2})$, where the subscript $k$ represents the $k^{th}$ decoherence probe, and $\Gamma_k$ is the coupling strength between the decoherence probe and the system, with $\Gamma_k = 10$ meV. The decoherence probes ($\Sigma_B$) are included at each nucleotide (the diagonal elements of $H_b$), except at the two contact locations. These are fictitious probes which mimic extracting electrons from the molecule and re-injecting them after phase breaking. The current at the $k^{th}$ probe is:

$$I_k^\uparrow = \frac{2q}{h} \sum_{l=1}^{N} \int_{-\infty}^{+\infty} T_{kl}(E)[f_k(E) - f_l(E)]dE = \int_{-\infty}^{+\infty} J_l(E)dE \quad (4)$$

where $k = 1,2,3,\ldots,N$, $T_{kl} = \Gamma_k G^r \Gamma_l G^a$ is the transmission function between probes $k$ and $l$, $G^a = (G^r)^\dagger$ is the advanced Green's function, $f_k(E) = \left(1 + \exp\left(\frac{E-E_{fk}}{kT}\right)\right)^{-1}$ is the Fermi distribution and $J_k(E)$ is the current with respect to energy at probe $k$. The condition for the fictious decoherence probes is that the current $J_k(E) = 0$ at every energy. These yields $N_b$ independent equations, where $N_b$ is the number of decoherence probes in the system. The following relation can then be derived [31]:

$$f_k(E) - f_L(E) = \left(\sum_{l=1}^{N_b} W_{kl}^{-1} T_{lR}\right)(f_R(E) - f_L(E)) \quad (5)$$

where $k = 1,2,3,\ldots,N$, $W_{kl}^{-1}$ is the inverse of $W_{ij} = (1 - R_{kk})\delta_{kl} - T_{kl}(1 - \delta_{kl})$, where $R_{kk}$ is the reflection probability at probe $k$, and is given by $R_{kk} = 1 - \sum_{k \neq l}^{N} T_{kl}$. Since the currents at the left $(I_L^\uparrow)$ and right $(I_R^\uparrow)$ are governed by the conservation of electron number, $I_L^\uparrow + I_R^\uparrow = 0$, this yields the equation for the current at the left contact [31], which is the current flowing through the DNA:

$$I_{DNA} = I_L^\uparrow = \frac{2e}{h} \int_{-\infty}^{+\infty} T_{eff}(E)[f_L(E) - f_R(E)]dE \quad (6)$$

The effective transmission term is:

$$T_{eff} = T_{LR} + \sum_{k=1}^{N_b} \sum_{l=1}^{N_b} T_{Lk} W_{kl}^{-1} T_{lR} \quad (7)$$

In equation (7), $T_{LR}$ is the coherent transmission from the left electrode to the right electrode. The second term is the decoherence contribution to transmission via the decoherence probes.

The final step involves the calculation of current under different biases and Fermi energies. The DNA structure is static throughout the calculations. We apply a potential shift in the DNA Hamiltonian to represent voltage drop across the DNA. In a metal-molecule-metal setup, obtaining the electrostatic potential profile across the molecule and finding the exact location of the Fermi energy requires using the density matrix and solving Poisson's equation in a self-consistent manner [32]. However, this is very difficult to achieve for large molecules such as the studied DNA (which consists of several hundred atoms). In addition, the details of the contacts such as the orientation and geometry at the metal-molecule interface are unknown, which further increases the complexity of the problem. Therefore, the general approach is to treat the Fermi energy as a variable and calculate the current or conductance at different Fermi energies. Further, at nonzero bias, a common approach is to use a linear ramp potential profile across the molecule by shifting the molecular orbitals in the DNA Hamiltonian ($H_b$) prior to calculating the transmission and the current. Following prior work [32], the potential drop is taken to be:

$$V_n = \begin{cases} 0, & n = 1 \\ qV_{bias} \times \left(0.4 + \dfrac{0.2(n-2)}{N-3}\right), & 2 \leq n \leq N-1 \\ qV_{bias}, & n = N \end{cases} \quad (8)$$

where $V_n$ is the voltage at base pair $n$. The electrostatic potential energy at base pair $n$, $qV_n$, is added to the diagonal elements of the Hamiltonian $H_b$ of block $n$. Note that block $n$ consists of a base pair, $q$ is the magnitude of electron charge ($1.6 \times 10^{-19}$ C), and $V_{bias}$ is the bias voltage applied across the strand. Locations 1 and $N$ represent the contact coupling sites. We set 40% of the voltage to drop on each end of the strand, while the remaining 20% of the bias drops linearly along the DNA.

The next step involves the calculation of current under different voltage biases and Fermi energies. The DNA structure is static in the calculations. After shifting the molecular orbitals in $H_b$, we calculate the effective transmission and use equation (6) to calculate the current at a certain bias with

$$E_{fR} = E_{fL} + qV_{bias} \quad (9)$$

where $E_{fL}$ and $E_{fR}$ are the Fermi energies of the left and right contacts, respectively.

Figure 2(a) shows the schematic for the DNA conductance calculations along with the non-linear potential drop, illustrating how the molecular orbitals shift along the strand. Figure 2(b,c) shows the transmission profiles of the two DNA strands (B-CT$_1$C and B-TC$_1$T) under different biases -1V (yellow), 0V (green), and 1V (blue). The transmission profile carries information regarding the density of states (DOS) of the system. Since DOS is directly related to the energy distribution of the molecular orbitals, the effect of shifting of the orbitals' energy due to the applied bias is observable in the transmission. The peaks in transmission suggest high DOS at those energies, which in turn yields high current conduction. From Figure 2(b,c), we can observe that the peaks shift to the right (higher energy) when a positive bias is applied and vice versa. Numerically, for B-CT$_1$C DNA, the HOMO peak is at -5.8 eV, -5.2 eV, and -4.78 eV corresponding to bias values of -1 V, 0 V, and 1 V respectively. Thus, we have a 10.38% and 8.47% shift in HOMO when the bias changes from 0 to -1 V and from 0 to 1 V respectively. Similarly, we get a shift of 8.6% and 9.4% in the case of B-TC$_1$T DNA when the bias changes from 0 to -1 V and from 0 to 1 V respectively. The transmission profiles at different energies and biases are plugged into equations (6) and (9), to get the transfer characteristics shown in Figure 2(d,e). The current-voltage curves for both DNA strands are shown at different Fermi energy values.

We have chosen to operate near the HOMO since the comparison between the gold work function and the ionization potentials of the DNA base pairs estimates the Fermi energy of gold to be close to the HOMO of the DNA. From Figure 2(d,e), we observe that there is a significant current difference between the two strands in the voltage range of 0 to 1 V. Therefore, we use these two strands to represent logic 0 and 1 memory states in our study of the DNA-ROM crossbar.

## 3. ARCHITECURE OF DNA CROSSBAR ARRAY

Traditional memory storage devices are fast reaching their scaling limits. Additionally, these devices suffer from low packing density of cells. To mitigate these problems, crossbar architectures are preferred for Resistive Random-Access Memory (RRAM) applications such as neuromorphic computing, in-memory computing, and image processing. Such an architecture is advantageous because of fast read and write operations which are essential for many digital and analog applications. Random access is enabled by selective application of a bias to appropriate wordlines and grounding of corresponding bitlines. For example, to access the memory element at the (1,3)[th] location in the array shown in Figure 3(a), a bias $V_1 = 1V$ has to be applied to wordline 1 with bitline 3 grounded (all other bitlines are kept floating). A high current measured in bitline 3 under this biasing scheme represents logic state '1' and a low current represents logic state '0.'

In this article, we propose and analyze the performance of a DNA crossbar array as a storage device. Figure 1(c) shows our proposed DNA ROM architecture. In this arrangement, a double-stranded (ds) DNA is assembled at every junction (alternately referred to as grid points or memory cell locations in this article) of a crossbar array. A circuit schematic of a DNA crossbar array with interconnect resistances and two different DNA strands, depicted as different resistors, is shown in Figure 3(a). Since the performance of the array depends significantly on the difference in conductance of the two dsDNA sequences, a proper choice of the DNA pair is critical.

Unlike an RRAM crossbar technology, DNA crossbars are not associated with a WRITE process which requires that resistances be updated by application of an external bias. Hence, we study the applicability of DNA in a ROM system instead of a RAM. In DNA-based storage devices, information can be easily loaded onto the device using sequence encoding. We use a readout scheme proposed by Liao et al. in [33]. In their scheme, to read the $i^{th}$ row, an external bias voltage $V_{in}$ is applied to that row (wordline) and the currents through the grounded columns (bitlines) are recorded. Extensive computational experiments have demonstrated that the readout approach outlined in [33] maintains high accuracy, while reducing the computational complexity from log($m^2n^2$) to log($m^2+n^2+mn$) for an $m\times n$ array. In performing our simulations, we have grounded all bitlines under the assumption that the information is stored in the whole crossbar.

The interconnect resistance $R_{int}$ of an RRAM crossbar between adjacent memory cells is about 7-10 Ω when the half-pitch is 14 nm. If the cross-sectional area of the interconnects is reduced in the interest of scalability, $R_{int}$ increases. However, large interconnect resistances (> 10 Ω) in RRAM arrays are problematic for READ and WRITE operations, especially the WRITE process, since the resistive state of the array cannot be set properly. Our proposed DNA crossbar is intended for data storage and retrieval applications. Therefore, it involves only a READ process. The resistance of DNA strands is typically on the order of mega-Ohms, which permits the use of higher interconnect resistances before sneak path issues start to adversely affect the performance of DNA crossbar arrays.

In this paper, we consider $R_{int}$ as purely a simulation parameter which models the interconnect resistance between two adjacent DNA memory cells and affects the accuracy and power consumption of the array. For DNA crossbar-based storage technology, potential candidates for interconnects include DNA nanowires (resistance of which is not yet known), carbon nanotubes (1-20 KΩ) [34], [35], and graphene nanoribbons (10-1000 KΩ) [36], [37]. Carbon nanotube interconnects can have a resistance of less than 10 kΩ depending on the number of shells. The resistance of DNA nanowires currently being pursued experimentally is expected to be larger than that of nanotubes. Conventional metal interconnects like tungsten and copper have a resistivity of $5.6 \times 10^{-8}$ Ω.m and $1.68 \times 10^{-8}$ Ω.m respectively. Accordingly, we have considered a wide range of $R_{int}$ values from 10 kΩ to 1 MΩ, without explicitly making a choice regarding the interconnect material or its dimensional aspects such as the cross-sectional area or aspect ratio. Our simulation results should provide valuable guidelines regarding the physical design parameters of a DNA crossbar array vis-à-vis its key performance parameters such as bit error rate and power dissipation.

In the subsequent sections, we analyze the performance of our proposed DNA crossbar array w.r.t bit error rate and power consumption due to READ operations only. We also demonstrate in detail how parameters such as the value of interconnect resistance and Fermi energy variations of the DNA strands from their HOMO levels can affect the performance of such an array.

## 4. ARRAY SIMULATION RESULTS

The detailed simulation methodology is shown in Figure 3(b). From DNA transport simulations at various voltage biases, a look-up table is constructed to calculate the DNA currents. Since the current-voltage characteristic of DNA strands is non-linear (see Figure 2), we have adopted an iterative approach for computing the voltage drop $V_{ij}$ across the DNA at the $(i,j)^{th}$ location of the memory array, after the interconnect resistances and the resistances of the other DNA memory elements in the array have been accounted for. In this paper, we compute $V_{ij}$ when a voltage of $V_{in} = 1\ V$ is applied to row $i$ while other rows are left floating, and the bottom of each column in Figure 3 is grounded. The iterative process commences with a small, applied voltage of 0.05 V to obtain the initial conductance matrix by interpolating from the look-up table. Based on the conductance values of DNA memory cells, the voltage distribution in the array is calculated using the following equation:

$$V_{ij} = \frac{1 + \sum_{k=j+1}^{n}\left(G_{ik} * \sum_{w=j+1}^{k}\frac{1}{g_{l_w}}\right)}{1 + \sum_{k=1}^{n}\left(G_{ik} * \sum_{w=j+1}^{k}\frac{1}{g_{l_w}}\right)} \times (V_{in}\alpha_i) \qquad (10)$$

where $i$ and $j$ are the row and column indices, $G_{ij}$ is the conductance of the DNA sequence at the $(i,j)^{th}$ grid point, $g_{l_w}$ is the interconnect conductance between adjacent memory cells along wordlines, and $\alpha_i$ is a row-specific sneak path parameter (see eqn. (8) in [33]) which is independent of the bias voltage. After the first iteration, the resistance states of DNA memory cells are updated according to the node voltages and the current-voltage relationship. The updated resistance states are then used to refine the voltage distribution. This process continues until the voltage difference between two successive iterations at each node is within an acceptable tolerance. The voltage distribution obtained after convergence is then used to compute the final read-out current distribution. We refer to the set of voltages, $\left\{\frac{V_{ij}}{V_{in}\alpha_i} : 1 \leq j \leq n\right\}$, as *normalized voltages*.

All simulations in this section were performed considering homogeneous arrays, which we define as arrays composed of identical DNA sequences at all junction points. We have modelled the crossbar array as a resistive network and each memory cell is modelled as a non-linear resistor. Capacitive effects, which are important in understanding the transient response, are not accounted for in our current study. While such effects can contribute to latency and play a critical role in matrix-vector multiplication applications, its impact on storage applications isn't major. To compute the latency, we need to calculate the quantum capacitance of DNA memory cells and interconnects. Carbon nanotubes offer a quantum capacitance of 100-400 aF/µm [38], [39]. Similar estimates for DNA origami nanowires have not been derived till date. Figure 4 shows the normalized voltages (applicable for all rows) for a homogeneous 64×64 array, composed with either B-CT$_1$C or B-TC$_1$T dsDNA sequences and $R_{int}$ = 10 kΩ or 1 MΩ. We chose a value of 10 kΩ on the low end to better demonstrate the sensitivities of array voltages (Figure 4) and currents (Figure 5) to the interconnect resistance.

Figure 4 also shows the normalized voltages for two different Fermi energies, at HOMO and (HOMO+0.2) eV. First, we observe that the voltage degradation is worse the farther a node is from the input, irrespective of the type of sequence and the value of $R_{int}$. Since the read current of a memory cell is directly proportional to the voltage across it, we expect this voltage degradation to be reflected in the spatial distribution of read currents, which in turn should affect the read accuracy. We address this issue in the subsequent paragraph. Second, we observe that the voltage degradation is worse for the B-CT$_1$C sequence compared to the B-TC$_1$T dsDNA sequence, for all values of interconnect resistance and Fermi energy. This is due to the presence of more low resistance cells in B-CT$_1$C crossbar configurations compared to B-TC$_1$T crossbar configurations, which leads to more sneak paths. From Figure 2(d,e), we see that the conductance of B-CT$_1$C dsDNA is higher than that of B-TC$_1$T strand, particularly when the applied bias is in the range of 0 to 1 V. At a particular interconnect resistance, voltage degradation is less at a higher value of Fermi energy for both the homogenous arrays. The reason for this can be attributed to the reduced conductance of the DNA strands at a higher Fermi energy (see Figure 2(d,e)).

From the preceding discussion, we conclude that the B-CT$_1$C dsDNA sequence is more susceptible to changes in the interconnect resistance due to its high conductivity. This is manifested in Figure 5 which shows the current distribution heatmaps for a 64×64 homogeneous array composed of *either* the B-CT$_1$C dsDNA sequence *or* the B-TC$_1$T dsDNA sequence, with $R_{\text{int}}$ = 10 kΩ and 1 MΩ and two different Fermi energies. The current $I_{ij}$ used to generate the 2-D heatmaps is the current that can be measured experimentally, i.e., it represents the current flowing out at the bottom of column *j* when a voltage of $V_{in} = 1 V$ is applied to row *i* while other rows are left floating. The expression for this current is:

$$I_{ij} = \beta_j * I_{DNA}(V_{ij}) \quad (11)$$

where $I_{DNA}(V_{ij})$ is the current flowing through the DNA at the (*i, j*)-th location of the memory array calculated using eqn. (6), $V_{ij}$ is the voltage drop across this DNA strand calculated using eqn.10, and *β$_j$* is a column-specific sneak path parameter (see eqn. (9) in [33] and Figure 3). Like $\alpha_i$, *β$_j$* is also independent of the bias voltage. Figure 6 shows the dependence of $\alpha_i$ and *β$_j$* on row and column indices respectively. Sequential application of the bias voltage to the rows of a crossbar array yields an output current matrix which is depicted as a heatmap in Figure 5.

Comparing the rows in Figure 5, we first observe that for both values of interconnect resistance, the spatial distortion of the read current distribution is lower for the B-TC$_1$T array compared to the B-CT$_1$C array. Second, we note that the distortion is greatest in the northeast corner of the array. This is due to the combined effect of the sneak path parameters, $\{\alpha_i, \beta_j: 1 \leq i \leq m, 1 \leq j \leq n\}$, which are plotted in Figure 6. It follows from Figure 6 that the largest distortion should occur when *i* = 1 (row) and *j* = *n* (column). As shown in [33], [40], these parameters are estimated from an approximation of Kirchhoff's laws (KCL and KVL) and is based on an average conductance of the memory cells. The results obtained from this parametric model with $\alpha$ and $\beta$ are in good agreement with the solution of Kirchhoff's equations, particularly for array sizes in the range 64×64 to 128×128 (see Figure 5(c) in [33]). By estimating the output current with $\alpha$ and $\beta$ parameters, this method provides remarkable improvement over a Kirchhoff's laws-based solution procedure in terms of time complexity, while maintaining almost similar accuracy. Finally, for both sequence types, we notice that the distortion in Figure 5 is significantly higher when $R_{\text{int}}$ = 1 MΩ. This can be attributed to an increase in the number of sneak paths and a higher voltage drop across the crossbar when the interconnect resistance is within one or two orders of magnitude of the resistance of the memory cell. Consequently, if binary information is stored in a crossbar that uses a B-CT$_1$C sequence to represent logic 0 and a B-TC$_1$T sequence to represent logic 1, we can expect some overlap between the current values recorded at the logic 1 locations and the logic 0 locations, leading to bit errors during the read process. Finally, we observe that the distortion is small when the Fermi energy of DNA strands is deep inside the bandgap. The reduced conductance at higher Fermi energy causes a smaller voltage to drop, which results in a smaller distortion.

## 4. IMPACT OF FERMI ENERGY VARIABILITY

Conductance dispersion in single-molecule systems can be attributed to several factors, including the geometry of atoms, the orientation of atoms at the interface, the bonding energy between molecule and contact, and stochastic rupture [21]. When a molecule touches a contact, partial charge transfer occurs due to hybridization which partially charges the molecule [41]. This in turn induces a change in the potential of the molecule. As a result, the molecular orbitals shift with respect to the Fermi energy of the contact, which we assume is fixed. We relate this orbital shift to the energy separation between the contact's Fermi energy and the HOMO level of the DNA. Ultimately, this translates to a Fermi energy variation of the DNA sequence. Since each dsDNA memory cell can be viewed as a single molecule system, the performance of the crossbar array depends on the Fermi energy variation of each DNA sequence as well as the spatial voltage/current distortion across the crossbar, as discussed in the previous section. In this section, we explore the extent to which Fermi energy variations can affect the performance of a DNA crossbar array. As before, we will use B-CT$_1$C and B-TC$_1$T dsDNA sequences to represent logic 0 and logic 1 respectively.

For a better understanding of the sensitivity of DNA currents to Fermi energy variations and their impact on crossbar performance, we have conducted 1000 simulations with a bias of 1 V, choosing a small random deviation of Fermi energy from the HOMO level in the range [0, $\delta$] for each simulation, where $\delta$ = 0.1 eV and 0.2 eV. The array size is 64×64 and the interconnect resistance, $R_{\text{int}}$, is chosen to be 100 kΩ or 1 MΩ.

The first row of Figure 7 shows the average voltage distribution for low ($\delta$ = 0.1 eV) and high ($\delta$ = 0.2 eV) Fermi energy variability. We observe that the mean voltage of the crossbar increases with an increase in Fermi energy variability $\delta$. Specifically, for $R_{\text{int}}$ = 100 kΩ and 1 MΩ, the mean crossbar voltage increases by 13.33% and 30.76% respectively when $\delta$ changes from 0.1 eV to 0.2 eV. From Figure 2(d, e), we observe that the conductance of both DNA strands decreases with an increase in $\delta$. This leads to a reduction in sneak paths in the crossbar architecture which translates to higher node voltages. For a fixed $\delta$, the mean crossbar voltage is higher for $R_{\text{int}}$ = 100 kΩ compared to 1 MΩ. This can be attributed to fewer sneak paths in the crossbar caused by lower interconnect resistance w.r.t. DNA resistance. Higher mean voltages are desirable since they translate to improved readout current distributions (reduced overlap) and enhanced read accuracy.

The second row in Figure 7 shows the distributions of read currents for a 64×64 array with different values of $\delta$ and $R_{\text{int}}$. For clarity, logic '1' currents are shown in blue, and logic '0' currents are shown in red (the probability of '1 cell' in the array is equal to the probability of '0 cells'). With $R_{\text{int}}$ = 100 kΩ, the overlap between bit 1 and bit 0 currents is very small for both low and high $\delta$. However, as the interconnect resistance increases to 1 MΩ, the current distributions overlap due to spatial distortion of current and sneak path effects, which implies that error-free reconstruction is not possible. Lower conductance of DNA strands with an increase in $\delta$ (see Figure 2(d,e)) and larger sneak path currents which are associated with higher interconnect resistance are the two main reasons for the overlap between the two class distributions.

In terms of bit-error rates (see the third row in Figure 7), we can see a small mean BER of 0.23% for $\delta$ = 0.1 eV and 1.48% for $\delta$ = 0.2 eV, when $R_{\text{int}}$ = 100 kΩ. The mean errors increase to 20.3% for $\delta$ = 0.1 eV and 25.8% for $\delta$ = 0.2 eV when $R_{\text{int}}$ = 1 MΩ. Thus, we see that Fermi energy variation plays a crucial role in determining the accuracy of the DNA crossbar as it causes an increase in mean BER by 1.25% and 5.2% at $R_{\text{int}}$ = 100 kΩ and 1 MΩ respectively. Even though we have not shown results for $R_{\text{int}}$ = 10 kΩ, we note that the distributions for 'bit 1' and 'bit 0' currents do not overlap, implying 0% BER (perfect readout), for both $\delta$ = 0.1 eV and 0.2 eV.

The preceding discussion leads to important observations. Since the optimal threshold level changes with $\delta$, ideally, one should determine it after fabrication of the array, based on the distribution of read currents. However, since the optimal threshold levels are in the order of nano-Amperes, the threshold detector must be accurate and precise enough to read and distinguish between current values of a similar or smaller order. This is the practical limitation which can affect the accuracies that may be achievable in practice.

## 5. EFFECT OF SCALABILITY ON IMAGE RETRIEVAL ACCURACY

In this section, we evaluate the performance of a DNA crossbar by loading 1000 randomly chosen images from the Linnaeus [42] dataset. Our evaluation criteria are BER and power consumption as a function of the bit load (in %), defined as the percentage of bits in an image that are represented by B-CT$_1$C DNA (logic '1'). Figure 8(a) shows the histogram of the bit load of all images in our dataset.

Figure 8 (b-g) show the bit error rate distributions as a function of bit load for different array sizes and interconnect resistances. The spatial bit distribution in a crossbar can be different for a particular bit load, leading to different bit errors. Therefore, we have presented box plots of BER for each bit load. As discussed in the previous sections, the bit error rate increases with an increase in array size and interconnect resistance. We observe that the BER is higher when the bit load is in the range of 50-70% (the critical range), and highest when it is approximately 60%. For reference, the BER is nearly 0% (in other words, the distributions of bit 1 and bit 0 currents do not overlap) for all interconnect resistances below 100 kΩ and 50 kΩ in cases of 64×64 and 128×128 arrays respectively while at 10kΩ the BER is exactly 0% for both the array sizes (see Figure 8(b,e)) resulting in perfect readout. Note that in Figures 4 and 5, we chose the interconnect resistance to be 10 kΩ and 1 MΩ in order to demonstrate the sensitivity of voltage and current distribution across the crossbar to extreme values of $R_{int}$. However, in Figures 7 and 8, we have chosen $R_{int}$ to be 100 kΩ and 1 MΩ since there are no bit errors when $R_{int}$ = 10 kΩ.

Figure 8 (h,i) show the average power consumed in the crossbar (memory cells and interconnects), as a function of $R_{int}$ and array size. With an increase in interconnect resistance, the mean voltage of the crossbar decreases, as seen from the trends in Figure 4 and Figure 7(a). This leads to lower read currents of the DNA strands. Both these factors collectively lead to lower power consumption, as can be observed in Figure 8 (h, i). For an array size of 64×64, the power consumed is 42.8 µW for $R_{int}$ = 10 kΩ, which reduces to 2.84 µW for $R_{int}$ = 1 MΩ. For an array size of 128×128, the power consumption is 60 µW for $R_{int}$ = 10 kΩ and 11.3 µW for $R_{int}$ = 1 MΩ. It follows therefore that a DNA array formed with a large interconnect resistance will scale better in terms of power consumption, but at the cost of an enhanced bit error rate.

## 6. CONCLUSION

To the best of our knowledge, this is the first paper of its kind to propose an electrically readable DNA storage device based on crossbar architecture. Our study leverages an idea that blends molecules created by synthetic biology with traditional semiconductor processes for contact formation, to yield a novel DNA storage technology. We have characterized the performance of DNA crossbar arrays as read-only memory devices in terms of readout accuracy and array power consumption. Since these performance metrics depend on the interconnect resistance, array size, and Fermi energy deviations of DNA strands, each of these parameters has been studied in detail. Our quantitative results provide a comprehensive benchmark for evaluating such (and competing) devices and shed important insights into how array size affects the trade-off between accuracy and power consumption. One important result from our analysis pertains to the role of interconnect resistance. We find that if the interconnect resistance is less than 10 kΩ on average, read errors are minimal for a 128×128 array. Second, the performance of a DNA-based crossbar array may be impacted by Fermi energy variations, which may be impossible to avoid completely in a single molecule system like DNA. Our study includes a comprehensive analysis of the impact of such variations on array performance. In our current analysis, we have studied the impact of different interconnect resistances and typical non-idealities in nanodevices which result in Fermi energy variability. Other parameters which should be considered in the future include contact engineering, non-uniformity of interconnects, and variability due to self-assembly process.

8. ACKNOWLEDGMENT

We acknowledge NSF Grant Numbers 1807391 (SemiSynBio Program) and 2036865 (Future of Manufacturing) for support. HM acknowledges Kuwait University Fellowship.

9. AUTHORS' CONTRIBUTION

MPA who works on DNA Nanotechnology defined the DNA memory framework. The transport calculation code was written by HM who also performed the calculations. The circuit modeling and image analysis were performed by AD. RK advised on circuit modeling and read technology. YW contributed to circuit-level data collection. The first draft of the manuscript was written by AD and supervised by AKD. All authors contributed to the subsequent draft of the paper and edited the manuscript.

10. DATA AVAILABILITY

All data are provided in this manuscript.

11. ADDITIONAL INFORMATION

All the authors declare no competing interests.

12. FIGURES & FIGURE LEGENDS

*Figure no. 1:*

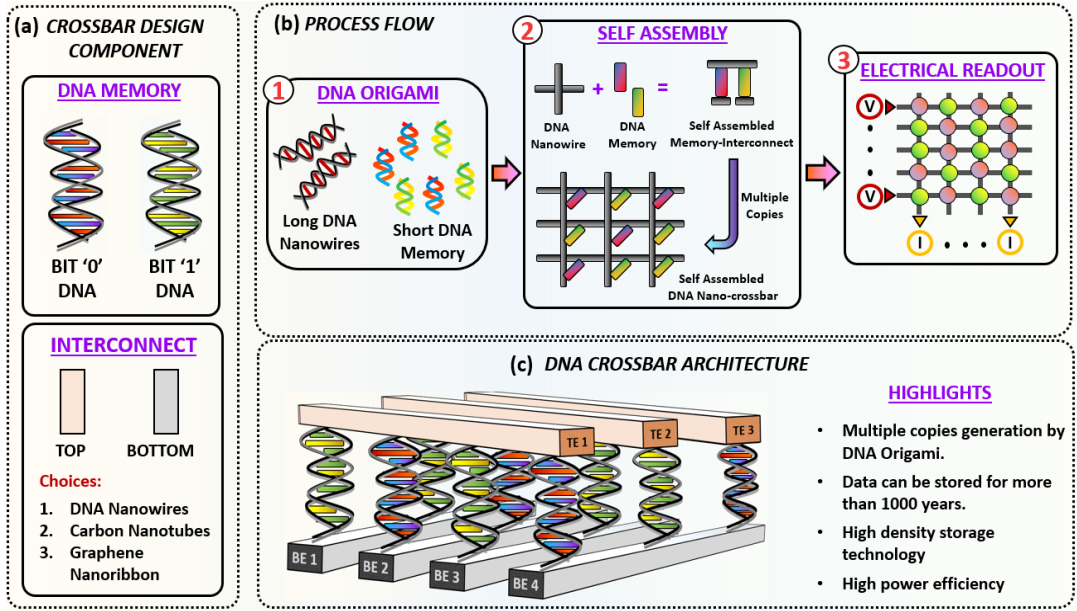

**Figure 1:** Illustration of a DNA based nano system for high-density data storage. (a) DNA crossbar components including memory cells and interconnects. (b) Process flow for the proposed technology. DNA origami principles can be invoked for nano-structure generation to form long DNA nanowires as interconnects and short DNA sequences as memory cells. A large number of these elements can be built through methods like PCR amplification. These components can then be self-assembled to create a crossbar architecture. An electrical read-out mechanism can be adopted for reading the information in the DNA crossbar array. (c) 3D conceptual schematic of a DNA crossbar array, where TE and BE denote the top and bottom electrodes respectively.

*Figure no. 2:*

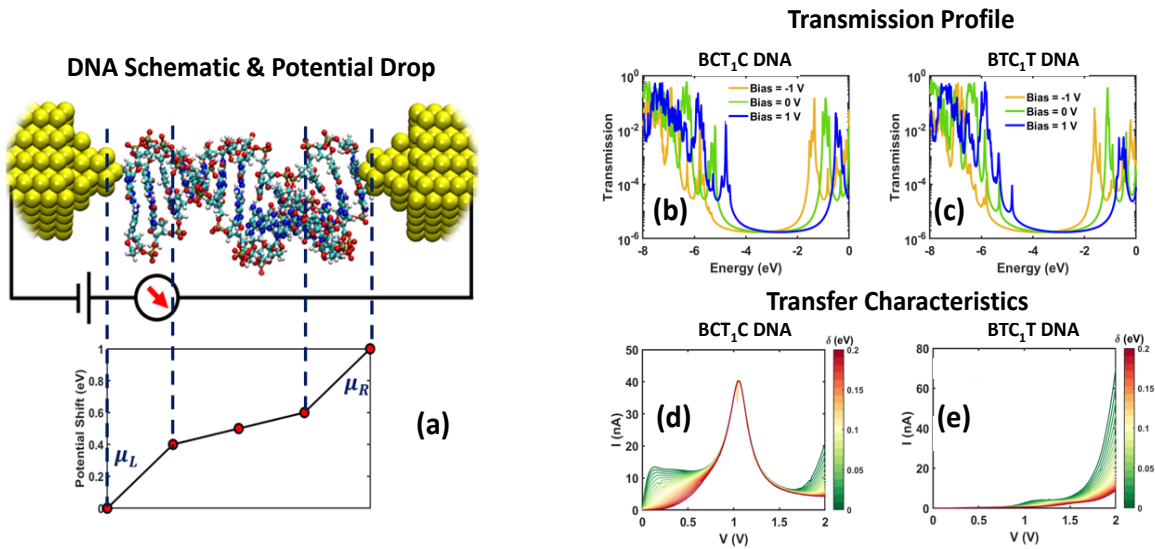

**Figure 2:** (a) A DNA strand is placed between two electrodes and bias is applied across it. For our work, B conformation of two double stranded DNA sequences is considered, $CT_1C$ and $TC_1T$. (b, c) Transmission versus energy for $CT_1C$ and $TC_1T$ strands as a function of Fermi energy for different bias voltages. (d, e) Current-voltage transfer characteristics for $CT_1C$ and $TC_1T$ B-DNA double stranded sequences. The collection of plots in these panels is for different Fermi energies ($\delta$).

*Figure no. 3:*

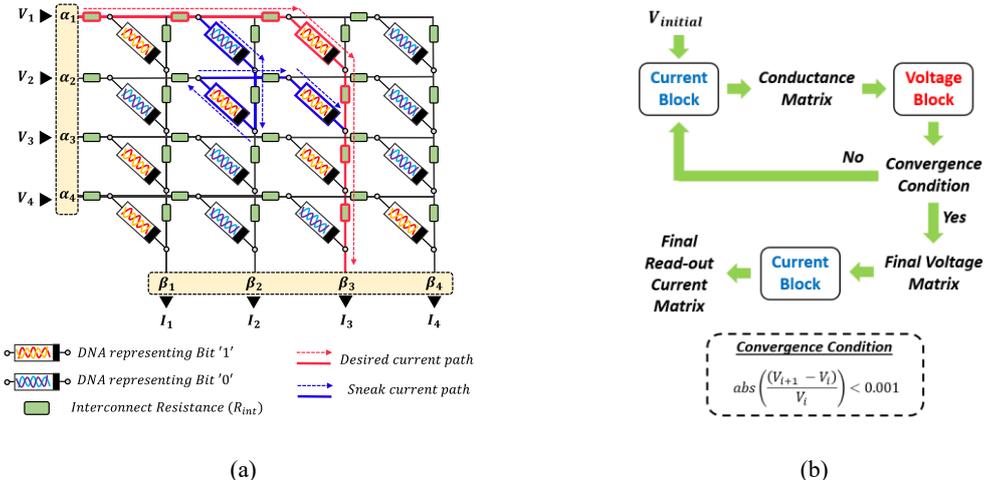

**Figure 3:** (a) Circuit schematic of a 4×4 array, where $\alpha_i$ and $\beta_j$ are array size dependent scaling factors which account for the interplay between interconnect resistance and sneak paths. (b) Flow chart for crossbar simulation. 'Voltage Block' comprises of the set of equations governing the voltage distribution in the crossbar (see eqn. 10) while 'Current Block' is a look-up-table created with the DNA currents ($I_{DNA}$) obtained from DNA current simulations (see eqn. 6).

*Figure no. 4:*

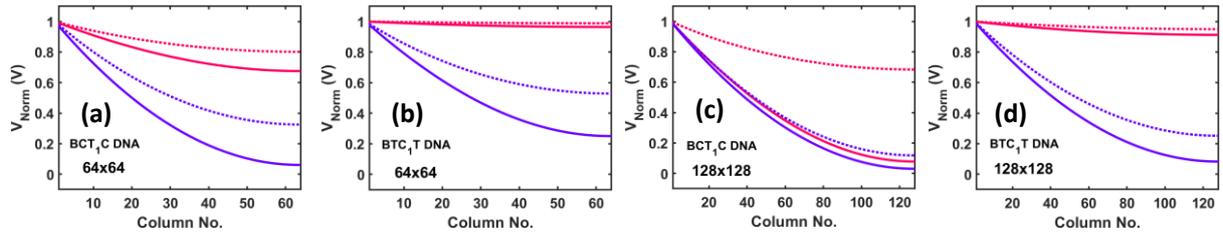

**Figure 4:** Normalized voltage distribution for homogenous arrays composed of B-CT$_1$C DNA (a, c) and B-TC$_1$T DNA (b, d). Array size is 64×64 for (a, b) and 128×128 for (c, d). The pink lines correspond to $R_{int}$ = 10 kΩ and the purple lines correspond to $R_{int}$ = 1 MΩ. Moreover, the solid lines represent Fermi energy at corresponding HOMO level while the dotted lines represent Fermi energy at (HOMO+0.2) eV.

*Figure no. 5:*

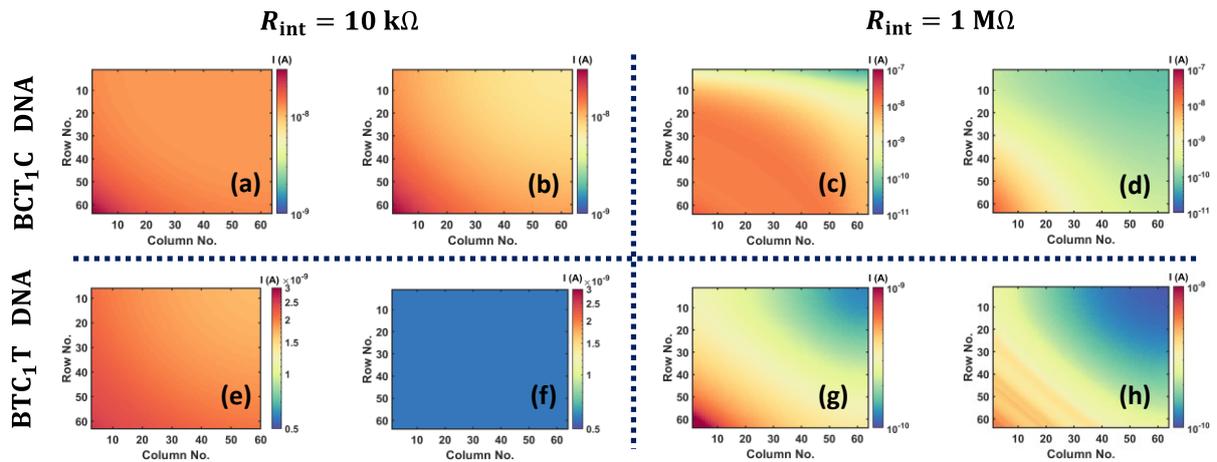

**Figure 5:** Read-out current distributions (see eqn. (11) for homogenous arrays composed of B-CT$_1$C DNA (a-d) and B-TC$_1$T DNA (e-h). The interconnect resistance is 10 kΩ for (a, b, e, f) and 1 MΩ for (c, d, g, h). Array size is kept fixed at 64×64. For panels (a, c, e, g), the Fermi energy is at corresponding HOMO level while for panels (b, d, f, h), Fermi energy is at (HOMO+0.2) eV.

*Figure no. 6:*

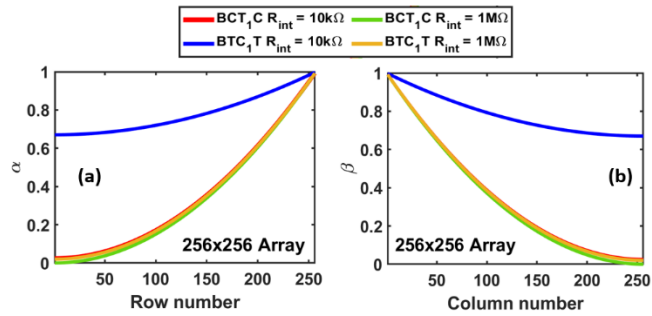

**Figure 6: (a) Plot of sneak path parameter *α* as a function of row index. (b) Plot of sneak path parameter *β* as a function of column index.**

*Figure no. 7:*

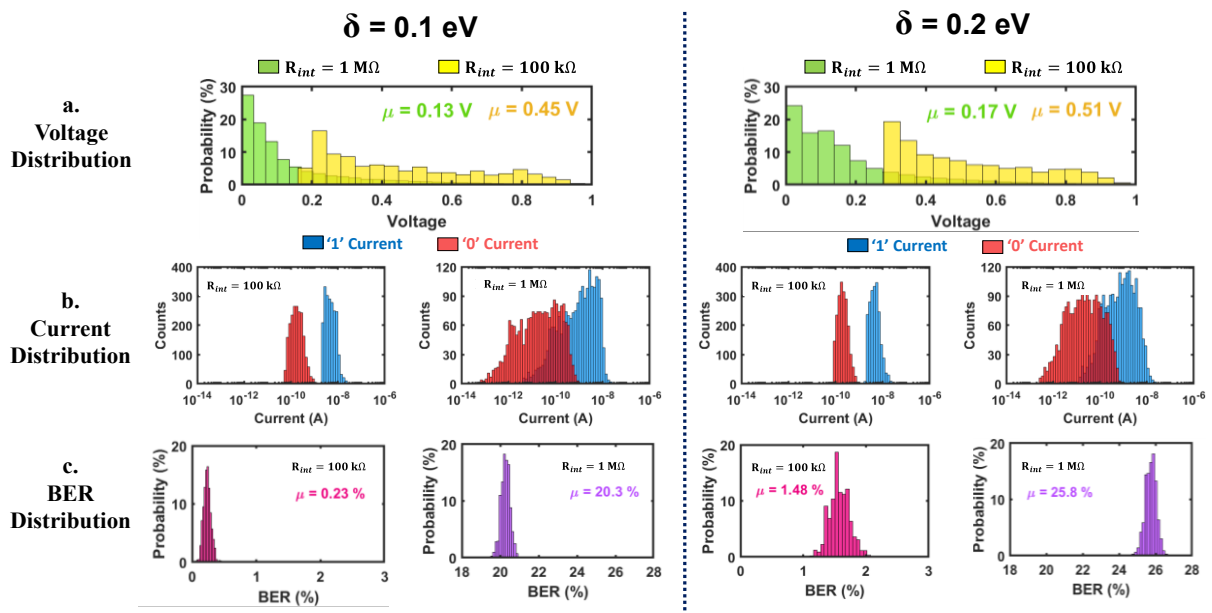

**Figure 7 Monte Carlo simulation results for a 64×64 array: (a) voltage distributions, (b) current distributions, (c) bit error rate (BER) distributions for low ($\delta$ = 0.1 eV) and high ($\delta$ = 0.2 eV) Fermi energy variations. All results (including the means) are computed from 1000 Monte Carlo simulations.**

*Figure no. 8:*

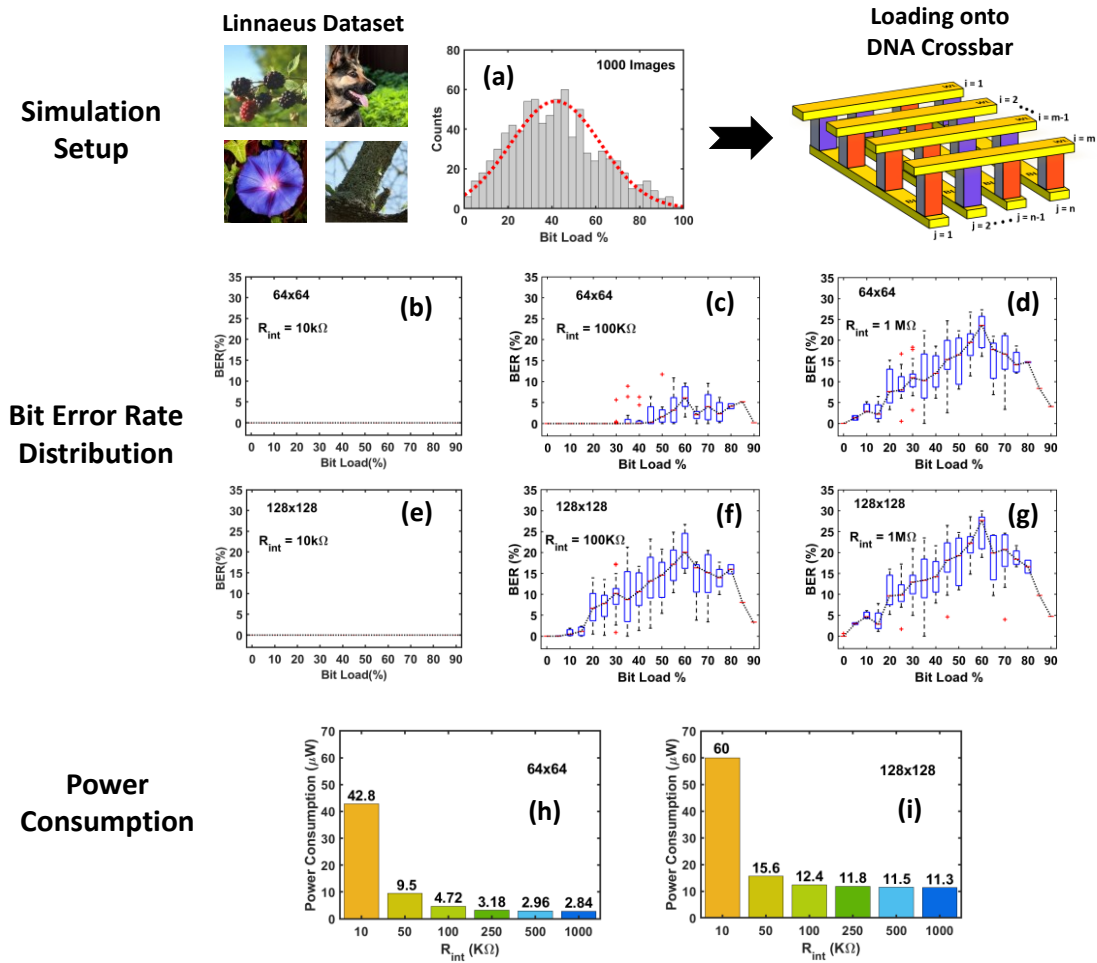

**Figure 8: Storage application results for a simulation setup where 1000 images from the Linnaeus dataset are loaded onto 64×64 and 128×128 crossbar arrays. (a) distribution of bit load (%) in the dataset, (b-g) distributions of bit error rates for each bit load (%) vs. array size and interconnect resistance, (h, i) average power consumed as a function of $R_{int}$ for 64×64 and 128×128 arrays. Note that the read-out accuracy is ~100% for interconnect resistances below 100 kΩ and 50 kΩ for 64×64 and 128×128 arrays respectively.**

TABLE 1

Comparison of various storage technologies (existing and proposed).

| Nature of Technology | 2D Density (bits/inch$^2$) | 3D Density (bits/inch$^3$) | Random** Access | Read$^{\$\$}$ Time | Lifetime (Years) |
|---|---|---|---|---|---|
| Conventional | $\sim 10^{12}$ [43] | $\sim 10^9$ [44] | Yes | Low | > 10 [45] |
| DNA-in-solution | – | $^{\#\#}6.04 \times 10^{22}$ [6] <br><br> $^{\#\#}1.07 \times 10^{23}$ [46] <br><br> $^{\#\#}1.18 \times 10^{24}$ [7] | No | High | > 100 (@ T = 25°C) <br><br> > 1000 (@ T < 10°C) [44] |
| RRAM crossbar | $4.5 \times 10^{12}$ [47] <br><br> $4.1 \times 10^{12}$ [48] <br><br> $4.2 \times 10^{12}$ [49] | – | Yes | Low | > 10 [45] |
| DNA$^{\&\&}$ crossbar (*This work*) | $3.47 \times 10^{12}$ (*12 nm*) <br> $4.72 \times 10^{12}$ (*10 nm*) <br> $6.78 \times 10^{12}$ (*8 nm*) | $7.35 \times 10^{18}$ (*12 nm*) <br> $1 \times 10^{19}$ (*10 nm*) <br> $1.44 \times 10^{19}$ (*8 nm*) | Yes | Low | > 100 (@ T = 25°C) <br><br> > 1000 (@ T < 10°C) [44] |

(##) For 3D density of DNA-in-solution technology, bulk density is that of pure water. ($$) Read time is classified to be 'High' if it is > 1 hour and 'Low' if it is < 10 ns. (&&) 2D and 3D density metrics for DNA crossbar storage are based on topological and DNA parameters. Italicized numbers (*8 nm*, *10 nm*, *12 nm*) refer to the half-pitch of the crossbar array.